\documentclass{emulateapj}
\usepackage{apjfonts}
\usepackage{lscape}
\input{epsf}

\slugcomment{To appear in the Astronomical Journal}

\shorttitle{Parallaxes of 18 nearby stars}
\shortauthors{L\'epine et al.}

\begin{document}

\title{New Neighbors: Parallaxes of 18 nearby stars selected from the
  LSPM-north catalog\altaffilmark{1}}

\author{S\'ebastien L\'epine\altaffilmark{2}, John R.
  Thorstensen\altaffilmark{3}, Michael M. Shara\altaffilmark{2}, \&
  R. Michael Rich\altaffilmark{4}}

\altaffiltext{1}{Based on observations obtained at the MDM
  Observatory, operated by Dartmouth College, Columbia University,
  Ohio State University, the University of Michigan, and Ohio
  University.}

\altaffiltext{2}{Department of Astrophysics, Division of Physical
  Sciences, American Museum of Natural History, Central Park West at
  79th Street, New York, NY 10024, USA; lepine,shara@amnh.org}

\altaffiltext{3}{Department of Physics and Astronomy, 6127 Wilder
  Laboratory, Dartmouth College, Hanover, NH 03755-3528;
  john.thorstensen@dartmouth.edu}

\altaffiltext{4}{Department of Astrophysics, University of California
  at Los Angeles, Los Angeles, CA 90095; rmr@astro.ucla.edu}

\begin{abstract}
We present astrometric parallaxes for 18 suspected nearby stars
selected from the LSPM-north proper motion catalog. Sixteen objects are
confirmed to be main sequence M dwarfs within 16 parsecs of the
Sun, including three stars (LSPM J0011+5908, LSPM J0330+5413, LSPM
J0510+2714) which lie just within the 10 parsec horizon. Two other
targets (LSPM J1817+1328, LSPM J2325+1403) are confirmed to be nearby
white dwarfs at distances of 14 and 22 parsecs, respectively. One of
our targets, the common proper motion pair LSPM J0405+7116E + LSPM
J0405+7116W, is revealed to be a triple system, with the western
component resolved into a pair of 16th magnitude stars (LSPM
J0405+7116W-A and LSPM J0405+7116W-B) with a 0.7$\pm$0.1$\arcsec$
angular separation. We find two stars (LSPM J1314+1320 and LSPM
J1757+7042) to be significantly overluminous for their colors, and
conclude that these may be unresolved doubles/multiples.
\end{abstract}

\keywords{astrometry \---  binaries: visual \--- stars: distances \---
  stars: low-mass, brown dwarfs \--- white dwarfs \--- solar
  neighborhood }

\section{Introduction}

The 20th century has seen considerable advances and efforts in
triangulating the distances of nearby stars through measurements of
their annual parallax. By 1995, over 8,000 stars had been monitored
using ground-based telescopes and their distances compiled in the Yale
Catalog of Trigonometric Parallaxes \citep{VLH95}. The Hipparcos 
mission has further increased this sample by over an order of
magnitudes, obtaining space-based astrometric parallaxes of over
110,000 stars \citep{P97}. These efforts have been fundamental to
modern astronomy, providing reliable absolute magnitudes for most
classes of stars, defining the first rung of the cosmic distance
ladder, constraining models of stellar structure and evolution, and
drawing a 3-D map of star systems in the vicinity of the Sun.

However, the map of the Solar Neighborhood remains fragmentary to this
day. The HIPPARCOS catalog lists 150 stars within 10 parsecs, and 1123
stars within 25 parsecs. However, the catalog is complete only to
about magnitude $V=8$, and reaches down only to about $V=12$. The local
stellar field is dominated by low-luminosity red dwarfs (main sequence
M dwarfs), which fall beyond the magnitude limit of the Hipparcos
catalog. As a comparison the Yale catalog lists 256 stars within 10
parsecs of the Sun, and 2059 stars within 25 parsecs, with most of the
additional stars consisting of low-luminosity red dwarfs and white
dwarfs. Apart from the low-luminosity companions of Hipparcos stars,
identified through common proper motion \citep{GC04,LB07}, astrometric
distances of low-luminosity objects, including brown dwarfs and most
red dwarfs and white dwarfs, are still largely dependent on
ground-based measurements.

\begin{deluxetable*}{lrrrl}[h]
\tabletypesize{\scriptsize}
\tablewidth{500pt}
\tablecolumns{5}
\tablecaption{Journal of Observations}
\tablehead{
\colhead{PM} & 
\colhead{$N_{\rm ref}$} & 
\colhead{$N_{\rm meas}$}  &
\colhead{$N_{\rm pix}$} &
\colhead{Epochs} \\
}
\startdata
J0011+5908  &   61  &  68  &  48  & 2005.71(6), 2005.88(12), 2006.64(9), 2006.84(13), 2007.73(8) \\
J0330+5413  &   55  &  87  &  64  & 2004.86(6), 2005.71(4), 2005.88(23), 2006.03(8), 2006.63(5), 2006.84(12), 2008.05(6) \\
J0336+3118  &   15  &  36  &  64  & 2005.88(11), 2006.04(1), 2006.66(3), 2006.83(14), 2007.08(5), 2007.72(16), 2008.04(6), 2008.14(8) \\
J0405+7116W  &   31  &  53  &  63  & 2004.86(7), 2005.20(6), 2005.71(5), 2005.87(13), 2006.04(2), 2006.66(6), 2006.84(10), 2007.07(8), 2008.05(6) \\
J0439+1615  &   28  &  61  &  57  & 2004.86(2), 2005.08(6), 2005.21(2), 2005.71(5), 2006.03(5), 2006.66(6), 2006.83(8),2007.72(13), 2008.04(6), \\
 & & & & 2008.15(4) \\
J0510+2714  &   46  &  97  &  71  & 2004.86(6), 2005.08(11), 2005.20(8), 2005.71(6), 2005.89(13), 2006.04(8), 2006.84(12), 2008.15(7) \\
J0515+5911  &   41  &  77  &  67  & 2004.86(6), 2005.08(11), 2005.21(5), 2005.71(7), 2005.88(12), 2006.03(10), 2006.84(10), 2008.05(6) \\
J0711+4329  &   24  &  29  &  56  & 2005.31(1), 2005.87(13), 2006.04(8), 2006.20(6), 2006.84(8), 2007.07(12), 2008.14(8) \\
J1119+4641  &    9  &  14  &  64  & 2005.09(10), 2005.20(7), 2005.30(10), 2006.05(5), 2006.20(6), 2006.38(6), 2007.08(3), 2007.34(8), 2007.91(5),\\
 & & & &  2008.14(4) \\
J1314+1320  &   14  &  20  &  53  & 2005.21(13), 2006.05(6), 2006.20(4), 2006.38(6), 2007.07(4), 2007.35(8), 2007.48(6), 2008.15(6) \\
J1428+1356  &   14  &  28  &  24  & 2005.21(3), 2006.39(8), 2007.34(3), 2007.47(6), 2008.15(4) \\
J1757+7042  &   32  &  58  &  64  & 2005.31(7), 2005.48(7), 2005.70(5), 2005.89(6), 2006.38(8), 2006.44(10), 2006.66(7), 2007.34(4), 2007.48(5),\\
 & & & &   2007.73(5) \\
J1817+1328  &   89  & 115  &  51  & 2005.48(6), 2005.70(6), 2006.38(7), 2006.44(10), 2007.34(8), 2007.48(6), 2007.72(8) \\
J1826+0146  &   33  &  49  &  79  & 2005.31(7), 2005.49(5), 2005.70(5), 2005.89(14), 2006.37(16), 2006.44(10), 2006.63(8), 2007.34(4), 2007.48(10) \\
J1839+2952  &   55  &  92  &  52  & 2005.30(10), 2005.48(7), 2005.70(5), 2006.38(1), 2006.44(10), 2006.63(1), 2007.35(6), 2007.47(9), 2007.73(3) \\
J1840+7240  &   36  &  58  &  73  & 2005.48(9), 2005.70(3), 2005.89(3), 2006.38(8), 2006.44(7), 2006.64(4), 2007.34(9), 2007.47(20), 2007.73(4),\\
 & & & &  2008.47(6) \\
J1926+2426  &  104  & 110  &  70  & 2005.49(3), 2005.70(6), 2005.88(8), 2006.37(9), 2006.44(19), 2006.64(10), 2006.84(8), 2007.34(7) \\
J2325+1403  &   18  &  43  &  46  & 2004.86(5), 2005.88(13), 2006.63(9), 2006.84(12), 2007.72(7) \\
\enddata
\tablecomments{Overview of the data included in the parallax solutions.
$N_{\rm ref}$ is the number
of reference stars used to define the plate solution, $N_{\rm meas}$
is the total number of stars measured, and  $N_{\rm pix}$ is the
number of images used.  The epochs represent different observing runs,
and the numbers in parentheses are the number of images included from
each run.}
\end{deluxetable*}

Parallax programs now in operation include the CTIOPI survey, carried
out from the SMARTS 1.5-meter and 0.9-meter telescopes in
Cerro-Tololo \citep{J05,C05,C06}. The survey includes several hundred
targets observable from CTIO, most of them high-proper motion stars
with photometric and spectroscopic distances placing them within 20
parsecs of the Sun, with an emphasis on very nearby (d$<$10 pc) objects
\citep{H06}. In the northern hemisphere, a smaller parallax program
has been carried on at the Allegheny observatory, and parallaxes of
21 nearby stars have recently been reported \citep{Gatewood09}. The
United States Naval Observatory has also been supporting a parallax
program in the past decades, from observations made at the USNO
Flagstaff station \citep{M92,Z00,Vrba04,Burgasser08,Dahn08}; the program has
notably contributed parallaxes for the first representative sample of
brown dwarfs \citep{D02}. The Torino Observatory Parallax Program has
also recently contributed data on six white dwarf candidates
\citep{smart2003} and 22 suspected nearby red dwarfs
\citep{smart2007}. Finally, parallaxes of selected nearby objects have
also been reported in recent years by various authors
\citep{Hambly1999,DH01,TK03,Deacon2005}, including parallaxes for 10
nearby T dwarfs \citep{TBK03}. Overall, it is a fair assessment that
these parallax programs are too modest in scope to keep up with the
large numbers of objects which have been identified as probable nearby
stars.

As of August 2008, the NStars database\footnote{http://nstars.nau.edu}
has parallax confirmation for 234 systems within 10 parsecs containing
325 individual stars, and for 2027 systems within 25 parsecs totaling
2629 stars. However, the census remains incomplete to this
date. Catalogs of stars with large proper motion still contain
thousands of stars which are suspected to be within the Solar
neighborhood but for which there is no parallax data. The large
catalogs assembled by W.J. Luyten through the 1970s, the LHS and NLTT
catalogs \citep{L79a,L79b} are still being mined for nearby star
candidates. Recent all-sky proper motion surveys have also added to
the bounty, identifying thousands more high proper motion stars that
had been overlook in earlier surveys
\citep{S02,Hambly2004,LS05,Subb05,L05,F07,L08}. The difficulty in
selecting nearby stars from proper motion catalogs is to separate the
stars whose large proper motion is due to their proximity, from the
stars whose large proper motion reflects a large transverse
velocity. For this, reliable photometric and/or spectroscopic
distances must first be obtained. A host of recent spectroscopic
follow-up observations have identified significant numbers of
candidate nearby stars
\citep{GR97,S01,S02,Scholz2004,S05,Henry2002,LRS03,Lodieu2005,RG05,Crifo05,PB06,Reyle06,Jahreiss2008}.
But the main break in extracting large samples of nearby stars from
proper motion catalogs has come from the availability of accurate
infrared photometry, from which reliable photometric distances can be
obtained from the optical-to-infrared color term
\citep{RC02,MSL02,Re02,H04,R04,RR04,L05,RCA07}. As a result, there are
now thousands of stars suspected to be within 25pc of the Sun but for
which there is no parallax distance confirmation. {\em The need for an
accurate map of the Solar vicinity thus justifies maintaining
existing nearby-star parallax programs, as well as the development
of new ones.}

Recently, \cite{T03} has demonstrated the use of the 2.4m Hiltner
telescope at MDM observatory to obtain reliable parallaxes at the 1-2
mas accuracy. MDM parallaxes of 6 known nearby stars were found to be
similar to USNO parallaxes of the same objects. Parallaxes of
27 relatively distant cataclysmic variables have so far been
successfully obtained from MDM \citep{T03,T06,T08}. Based on this
demonstrated success of using the Hiltner telescope for accurate
astrometry, we have initiated an program to measure the parallaxes of
selected nearby star candidates from the list generated by
\citet{L05}. 

In this paper, we present the first results of our program, which
yield the first parallax determinations for 18 stars predicted to be
within 15 parsecs of the Sun. Our astrometric observations are
described in \S2. We analyze the results in \S3, where we also test
the accuracy of the photometric distance estimates for low-mass
stars. Conclusions follow in \S4.

\section{Observations}

\subsection{Target selection}

We have observed a subsample of sixteen suspected nearby main sequence
M dwarfs, selected from our own list of candidates from the
LSPM-north proper motion catalog \citep{LS05}. The stars all have
photometric distances placing them within 15 parsecs of the Sun
\citep{L05}. Several of them have been proposed to be nearby stars in
recent years, based on photometric/spectroscopic distance estimates
from various authors (see \S 2.6).

We have also selected from the LSPM-north two new candidate white
dwarfs, both suspected to be within 12 parsecs of the Sun based on
photometry alone. Their presumed status as white dwarfs is strongly
suggested by their location in the reduced proper diagram, where they
show up as blue stars of low luminosity.

\begin{figure*}[t]
\epsscale{1.0}
\plotone{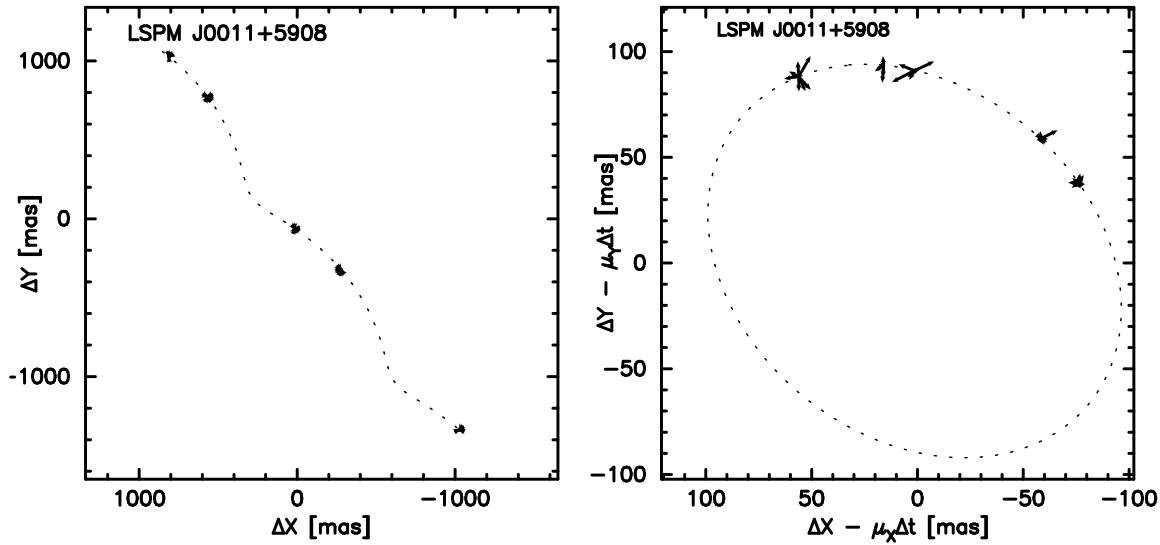}
\caption{Astrometric solution for the star LSPM J0011+5908 based on our
  five-epoch measurements. Left panel: motion on the plane of the sky,
  with the dashed line showing the solution for the combined proper
  motion and parallax. Right panel: fit of the parallactic ellipse,
  after the proper motion has been subtracted. Each vector shows the
  residual from the fit for each astrometric measurement; each epoch
  having several data points. Note that the shape and orientation of
  the parallactic ellipse is pre-constrained by the star's position on
  the sky and is not part of the fit; only the size of the ellipse
  is fit to the data points and yields the parallax measurement.}
\end{figure*}

\subsection{Astrometry and photometry}

Astrometric imaging was carried out from the 2.4-meter Hiltner
telescope at MDM observatory, located on the southwest ridge of Kitt
Peak, Arizona. Observations were scheduled in blocks of 4-6 nights,
separated by 1-3 months over a period of 3 years. The general program
included observations of several classes of targets, including
cataclysmic variables, low-mass halo subdwarfs, and our test sample of
18 suspected very nearby stars.

Observations were made with the MDM $2K\times2K$ CCD camera (dubbed
``Echelle''). A typical pointing consisted in a series of 5-12 short,
30-60 seconds exposures using either of the I band filter or a
$7000$\AA\ narrowband filter, which was used for targets bright enough
to saturate the camera in a 30s I-band exposure. An additional
exposure in the V band was typically obtained in each visit. Stars
were observed within 1 hour of the meridian whenever possible, but
sometimes as far as 2 hours. A journal of observations for the 18
nearby star candidates is provided in Table 1 for reference.

Astrometric solutions were determined following the method described
and documented in detail in \cite{T03}, including the setup of
reference objects in the field, corrections for field scale and
orientation, and a correction for differential atmospheric
refraction. A typical astrometric solution for one of the stars (LSPM
J0011+5908) is displayed in Figure 1. The vector root-mean-square
centroiding accuracy for well-exposed stars was typically 5 to 7 mas
per exposure, as judged from the scatter around the best fits; these
residuals are shown in Figure 1 for LSPM J0011+5908. We estimated the
parallax errors both from the formal fit errors and from the scatter
of the fitted parallaxes of the reference stars.

From our extended sets of CCD images, we also calculated apparent
magnitudes in the V and I bands. Photometry was calibrated from
observations of a set of photometric standards \citep{L92}. Most of
our fields have photometry from three or more observing runs, with
independently derived calibrations. Comparisons between runs indicate
that the zero points for the quoted magnitudes and colors should be
accurate to $<0.05$ mag.  Although we observed standard stars that
spanned a wide range of color (0.0$<$V-I$<$2.0), most of the program
stars are so red (V-I$>$3) that they required some extrapolation of
the color transformation. This can lead to systematic effects that
are difficult to estimate reliably, but the good run-to-run
reproducibility suggests that the $V-I$ colors of the reddest
objects are determined to better than 0.2 mag. Fortunately, 2MASS
infrared magnitudes are available for our program objects; the
$V-J$ colors are relatively insensitive to small errors in $V$, and
the latter were used to estimate the photometric distance. Table 2
present our final reduced astrometric and photometric results.

\begin{deluxetable*}{lcrrrrrrrrrrrrr}[h]
\tabletypesize{\scriptsize}
\tablecolumns{15} 
\tablewidth{520pt} 
\tablecaption{Astrometric and photometric data} 
\tablehead{
\colhead{LSPM star} &
\colhead{other name} &
\colhead{$\alpha$} &
\colhead{$\delta$} &
\colhead{$V_e$\tablenotemark{1}} &
\colhead{$\pi_{\rm rel}$} &
\colhead{$\mu_{\alpha}$} &
\colhead{$\mu_{\delta}$} &
\colhead{$\pi_{\rm abs}$} &
\colhead{$V$} &
\colhead{$V-I$} &
\colhead{$V-J$\tablenotemark{2}} &
\colhead{$M_V$} &
\colhead{dist.} \\
\colhead{} &
\colhead{} &
\colhead{(ICRS)} &
\colhead{(ICRS)} &
\colhead{mag} &
\colhead{mas} &
\colhead{mas yr$^{-1}$} &
\colhead{mas yr$^{-1}$} &
\colhead{mas} &
\colhead{mag} &
\colhead{mag} &
\colhead{mag} &
\colhead{mag} &
\colhead{pc} \\
}
\startdata
J0011+5908 & \nodata    &00 11 31.81& +59 08 39.9& 15.87& 106.5(0.9) & -901& -1167 & 108.3(1.4) & 15.68 & 3.67 & 5.74& 15.86 &  9.23$\pm$0.12\\    
J0330+5413 & \nodata    &03 30 48.89& +54 13 55.1& 16.20& 101.8(1.4) & -150&    -5 & 103.8(1.4) & 15.96 & 3.90 & 5.79& 16.04 &  9.63$\pm$0.13\\    
J0336+3118 & \nodata    &03 36 08.70& +31 18 39.6& 14.52&  78.0(2.5) &  120&  -125 &  79.6(2.5) & 13.72 & 2.58 & 4.53& 13.22 & 12.56$\pm$0.39\\    
J0405+7116E& G 221-27   &04 05 57.50& +71 16 40.8& 14.04&  55.7(1.1) &  172&  -379 &  57.1(1.2) & 14.01 & 2.87 & 4.48& 12.79 & 17.51$\pm$0.37\\    
J0405+7116W-A& LP 31-302&04 05 56.55& +71 16 38.2& 16.02&    \nodata &  172&  -379 &  57.1(1.2) & 16.03 & 3.38 & 5.93& 14.82 & 17.51$\pm$0.37\\
J0405+7116W-B& \nodata  &    \nodata&     \nodata&      &    \nodata &  172&  -379 &  57.1(1.2) & 16.88 & 3.79 &\nodata & 15.66 & 17.51$\pm$0.37\\
J0439+1615 & LHS 1690   &04 39 31.63& +16 15 43.0& 16.24&  85.0(2.0) &  -79&  -796 &  86.6(2.5) & 15.74 & 3.60 & 5.60& 15.43 & 11.55$\pm$0.33\\    
J0510+2714 & \nodata    &05 10 20.09& +27 14 01.9& 17.80&  98.9(1.2) & -214&  -631 & 100.7(1.6) & 17.29 & 3.90 & 6.59& 17.30 &  9.93$\pm$0.16\\    
J0515+5911 & \nodata    &05 15 30.96& +59 11 17.5& 18.44&  63.9(1.3) &  112& -1003 &  65.7(1.3) & 18.11 & 4.41 & 6.79& 17.20 & 15.22$\pm$0.30\\    
J0711+4329 & LHS 1901   &07 11 11.44& +43 29 58.0& 15.87&  76.5(1.5) &  353&  -576 &  77.8(3.0) & 15.90 & 3.98 & 5.92& 15.36 & 12.85$\pm$0.50\\    
J1119+4641 & LHS 2395   &11 19 30.61& +46 41 43.2& 16.30&  95.6(2.6) &  314&  -601 &  97.0(2.6) & 15.86 & 3.78 & 5.77& 15.79 & 10.31$\pm$0.28\\    
J1314+1320 & NLTT 33370 &13 14 20.39& +13 20 01.2& 15.93&  59.8(2.8) & -243&  -186 &  61.0(2.8) & 15.83 & 3.89 & 6.08& 14.76 & 16.39$\pm$0.75\\    
J1428+1356 & LHS 2919   &14 28 04.17& +13 56 13.3& 18.31&  81.9(4.1) & -359&  -475 &  82.8(4.1) & 17.65 & 4.30 & 6.64& 17.24 & 12.08$\pm$0.60\\    
J1757+7042 & LP 44-162  &17 57 15.40& +70 42 01.4& 18.84&  51.5(1.1) &   13&   328 &  52.4(1.1) & 18.79 & 4.67 & 7.34& 17.38 & 19.08$\pm$0.40\\    
J1817+1328 & \nodata    &18 17 06.50& +13 28 25.0& 16.05&  69.7(0.8) & -437& -1110 &  70.3(1.2) & 15.90 & 0.98 & 1.52& 15.13 & 14.22$\pm$0.24\\    
J1826+0146 & NLTT 46476 &18 26 16.56& +01 46 20.9& 15.91&  51.5(1.6) & -144&   314 &  53.5(2.0) & 14.94 & 3.27 & 4.90& 13.59 & 18.69$\pm$0.70\\    
J1839+2952 & LP 335-12  &18 39 33.08& +29 52 16.5& 17.93&  78.3(1.2) &   80&  -219 &  79.3(2.0) & 18.21 & 4.61 & 7.20& 17.70 & 12.61$\pm$0.32\\    
J1840+7240 & LP 44-334  &18 40 02.39& +72 40 54.1& 17.67&  58.4(2.0) &  -33&   189 &  59.3(2.2) & 17.62 & 4.37 & 6.65& 16.48 & 16.86$\pm$0.63\\    
J1926+2426 & G 185-23   &19 26 01.62& +24 26 17.2& 15.10&  51.3(0.9) &  182&   104 &  52.8(1.5) & 14.30 & 3.11 & 4.68& 12.91 & 18.94$\pm$0.54\\    
J2325+1403 & LP 522-46  &23 25 19.87& +14 03 39.5& 15.84&  43.6(1.6) &  345&   122 &  44.9(2.0) & 16.05 & 0.95 & 1.54& 14.32 & 22.27$\pm$0.99\\ 
 \enddata
\tablenotetext{1}{Visual magnitude estimated based on the photographic $b$, $r$, and $i$ magnitude
  from the Palomar Sky Survey plates \citep{LS05}.}
\tablenotetext{2}{J band magnitude from the 2MASS All-Sky Catalog on
  point sources \citep{2MASS}.}
\end{deluxetable*}

\subsection{LSPM J0405+7116: a triple system}

Our target list included one known double star: the common proper
motion pair composed of the V=14 star LSPM J0405+7116E, and its
companion the V=16 star LSPM J0405+7116W. The two stars have an
angular separation of $5\arcsec$ on the sky. The system made it into
our sample of candidate nearby stars because of the short photometric
distance of its secondary, which places it at a distance of only
12.6$\pm0.7$ parsecs \citep{L05}. The primary's photometry places it
at a somewhat longer distance of 14.6$\pm1.2$ parsecs. 

Our astrometric images revealed that, in fact, LSPM J0405+7116W is
itself a visual double, with the two components (LSPM J0405+7116W-A
and J0405+7116W-B) separated by $\approx0.7\arcsec$. The system is
thus revealed to be a triple. Our best-seeing images (0.9\arcsec) from
October 2006 shows the two stars just barely resolved, with an angular
separation $\rho$=0.7$\pm$0.1\arcsec. The south-west component, which
is marginally fainter, makes a position angle
pma=$245^{\circ}\pm10^{\circ}$ on the sky relative to the north-east
component. We measure the centroid of the two components in turn to be
5.3$\pm$0.1\arcsec away from G 221-27, with a position angle
pma=$240^{\circ}\pm4^{\circ}$. 

The three stars are listed in Table 2, with the two ``west''
components tabulated both individually and as a pair. While we have V
and I magnitudes for W-A and W-B, we only have 2MASS J magnitudes for
the unresolved pair.

\subsection{Spectroscopy}

Medium-resolution spectra were obtained for 11 of the targets as part
of our ongoing spectroscopic follow-up survey of stars from the
LSPM-north proper motion catalogue. Four stars were observed at
the Lick observatory with the 3-meter Shane telescope equipped with
the KAST dual-channel spectrograph. Three more stars were observed at
MDM on the 2.4-meter Hiltner telescope with the MkIII
spectrograph. The other four stars were observed at MDM on the
1.3-meter McGraw-Hill telescope, also with the MkIII
spectrograph. Standard reduction of all the spectra was performed with
IRAF. Spectra were all wavelength calibrated against NeAr comparison
arcs, and flux calibrated based on observations of the NOAO
spectrophotometric standards Feige 66, Feige 67, and Feige
110. Spectra from three of the stars (LSPM J0011+5908, LSPM
J0510+2714, and LSPM J1817+1328) were published an earlier paper
\citep{LRS03} but are presented here again for completeness.

Nine of the stars are confirmed to be late-type M dwarfs, with
spectral subtypes between M4.5 and M8.0. Spectral subtypes are
determined based of the strength of the CaH, TiO, and VO molecular
bands, as described in \citet{LRS03}. The tenth object, LSPM
J1817+1328, is confirmed to be a cool DA white dwarf with a very
weak, but detectable, H$\alpha$ absorption line. A blackbody fit of
the spectral energy distribution indicates a spectral subtype of DA
10. The reduced spectra are displayed in Figure 2.

\begin{figure}
\epsscale{2.2}
\plotone{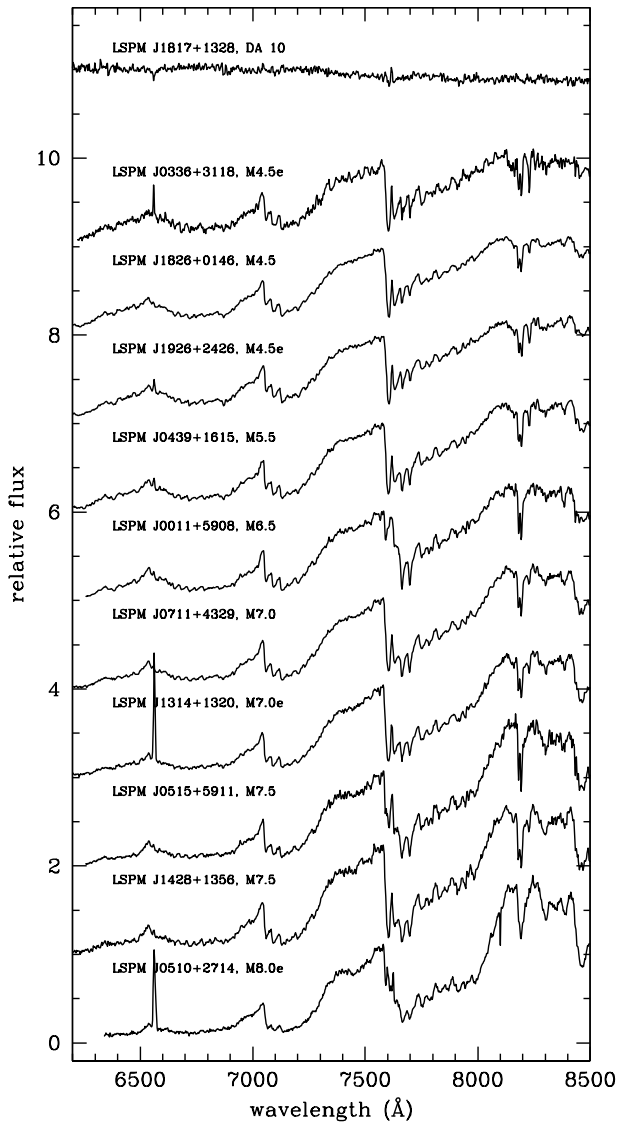}
\caption{Medium-resolution spectra of ten of our parallax
  targets, with names and spectral subtypes noted. Spectra are
  normalized to 1.0 at 8,100\AA, and shifted vertically by integer
  units for clarity. Except for the DA 10 white dwarf LSPM J1817+1328,
  all stars are mid-type to late-type M dwarfs, which is consistent
  with them all being nearby low-mass stars.}
\end{figure}

We searched the literature for published spectroscopic data on the
other 7 systems, and found formal spectral classifications for five of
them, including spectral subtypes for LSPM J0405+7116E and for the
the (unresolved) pair LSPM J0405+7116W-A and LSPM J0405+7116W-B
\citep{CR02}. All available spectral subtypes are tabulated in Table
3, along with the bibliographical source for the classification.

\begin{deluxetable}{lcccc}
\tabletypesize{\scriptsize}
\tablecolumns{6} 
\tablewidth{240pt} 
\tablecaption{Spectral classification and activity} 
\tablehead{ LSPM \# & sp.type & Ref.\tablenotemark{1} & H$\alpha$ EW &  X-ray\tablenotemark{2} \\
                    & & & \AA & counts s$^{-1}$}
\startdata 
J0011+5908 &    M6.5 &    L09  &    -0.9& \nodata\\
J0330+5413 & \nodata & \nodata & \nodata&  0.0188\\ 
J0336+3118 &    M4.5e&    L09  &    -5.3&  0.0737\\
J0405+7116E&    M4.0 &    CR02 & \nodata& \nodata\\
J0405+7116W(A+B)& M5.0e&  CR02 &    -8.9&  0.0970\\
J0439+1615 &    M5.5e&    L09  &    -2.6& \nodata\\
J0510+2714 &    M8.0e&    L09  &   -49.6&  0.0648\\
J0515+5911 &    M7.5 &    L09  &    -0.8& \nodata\\
J0711+4329 &    M7.0 &    L09  &    -0.7& \nodata\\
J1119+4641 &    M5.5 &    GR97 & \nodata& \nodata\\
J1314+1320 &    M7.0e&    L09  &   -54.1& \nodata\\
J1428+1356 &    M7.5e&    L09  &    -1.6& \nodata\\
J1757+7042 &    M7.5 &    G00  &    -0.9\tablenotemark{3}& \nodata\\
J1817+1328 &    DA11 &    L09  & \nodata& \nodata\\
J1826+0146 &    M4.5 &    L09  &    -0.4& \nodata\\
J1839+2952 &    M6.5 &    R03  & \nodata& \nodata\\
J1840+7240 &    M6.5 &    R04  & \nodata& \nodata\\
J1926+2426 &    M4.5e&    L09  &    -2.7&  0.0194\\
J2325+1403 &    DA10 &    VK03 & \nodata& \nodata
 \enddata
\tablenotetext{1}{Source of the spectral classification: L09 = this
  paper, CR02 = \citet{CR02}, GR97 = \citet{GR97}, G00 = \citet{G00},
  R03 = \citet{R03}, R04 = \citet{R04}, VK03 = \citet{VK03}}
\tablenotetext{2}{X-ray count rate from the ROSAT All-Sky catalogs
  \citep{V99,V00}.}
\tablenotetext{3}{H$\alpha$ equivalent width from \citet{schmidt2007}.}
\end{deluxetable}

\subsection{X-ray emission and activity}

We have examined our spectra for signs of activity, which in M
dwarfs is diagnosed by a strong H$\alpha$ line in emission. H$\alpha$
emission was detected in 9 of out targets; equivalent widths are
listed in Table 3. Four of the stars have H$\alpha$ equivalent widths
in excess of 2\AA, which qualifies them as active stars.

We further searched for bright X-ray counterparts in the ROSAT All-Sky
catalog of point sources \citep{V99} and ROSAT All-Sky catalog of
faint sources \citep{V00}. X-ray sources within 20 arcseconds of our
targets were selected as probable counterparts. The search turned up
probable X-ray counterparts to five of our target M dwarfs; count
rates per second are noted in Table 3. One of the stars for which we
lack spectroscopic data happens to be an X-ray bright source. The
ROSAT X-ray count rate for LSPM J0011+5908 is at a level similar to
the H$\alpha$-bright stars. We take this as an indication that LSPM
J0011+5908 is also an active M dwarf. We predict that spectroscopic
observations should reveal the presence of significant $H\alpha$
emission in that object.

Overall, most of our active objects show only moderate signs of
activity, typical of M dwarfs with detected H$\alpha$ emission. The
only exception is LSPM J0510+2714, whose H$\alpha$ emission is quite
significant.

\subsection{Notes on individual objects}

\subsubsection{LSPM J0011+5908}

First identified as a high proper motion star by \citet{LSR02}, the
very large proper motion ($\mu=1.48\arcsec$ yr$^{-1}$) was confirmed
by \citet{Levine2005}. The star was classified as M5.5 by
\citet{LRS03}, who estimated a spectroscopic distance of
12$\pm$4pc. Our own spectrum suggests a spectral type of M6.5 with a
weak but detectable H$\alpha$ emission. A photometric distance of
11.7$\pm$3.4pc was estimated by \citet{L05}. Our parallax places the
star safely within the 10 pc horizon, at a distance of
9.23$\pm$0.12pc. The large proper motion is consistent with a
transverse velocity $v_T$=64.9$\pm$0.8 km s$^{-1}$ which suggest it
could be from the old Galactic disk population. This would be
consistent with the apparent low chromospheric activity.

\subsubsection{LSPM J0330+5413}

This one is a very recent discovery, identified as a high proper
motion star by \citet{LS05}. With a proper motion of only
$\mu = 0.151\arcsec$ yr$^{-1}$, the star does not particularly stand
out among the nearby stars, which tend to have much larger proper
motions; the star however was estimated to be at 12.5$\pm$4.2 based on
photometry. While the star has no formal spectral classification as
yet, its color suggests a spectral type of about M5. Our geometric
parallax again places it within the 10 parsecs horizon, at a distance
d=9.63$\pm$0.13pc. The proper motion is consistent with a relatively
low transverse velocity of 6.9$km$ s$^{-1}$, which suggests the star
may be part of the young Galactic disk population. The moderate levels
of X-ray would be consistent with this suggestion; it would be
interesting to verify whether the star also shows significant H$\alpha$
emission.

\subsubsection{LSPM J0336+3118}

This star was also first identified as a high proper motion star in
\citet{LS05}. A photometric distance of 10.9$\pm$3.8pc was estimated
by \citet{L05}. The star happens to be in the direction of the Perseus
star forming  region. Its polarization was measured in a study of the
Perseus dark cloud complex \citep{goodman90} but found to be
negligible, consistent with the star being a foreground object. Our
parallax indeed places the star at a distance of only
12.56$\pm$0.39pc. The relatively low proper motion yields a transverse
velocity of only 10.6$\pm$0.3 km s$^{-1}$ consistent with the young
disk population. The star indeed shows signs of being relatively
young, with significant H$\alpha$ and X-ray emission. We classify the
star as M4.5e.

\subsubsection{LSPM J0405+7116E = G 221-27}

One of the high proper motion objects from the Lowell proper motion
survey \citep{GBT71}; its large proper motion was re-measured and
updated by \citet{SG03} and \citet{LS05}. The M4.0 dwarf was first
suspected to be a very nearby star by \citet{CR02} based on a spectroscopic
distance estimate of 14.7$\pm$1.2pc. \citet{S05} however
estimated a spectroscopic distance of 17.0pc with a $\pm$20\%
uncertainty, while \citet{L05} estimated a photometric distance of
19.4$\pm$7.2pc. Our parallax places the star at 17.51$\pm$0.37pc, more
in line with the Scholz estimates. The large proper motion yields a
transverse velocity of 34.9$\pm$0.7 km s$^{-1}$. The star is found to
be the more massive component in a triple system.

\subsubsection{LSPM J0405+7116W-A = LP 31-302}

This companion to G 221-27 was initially identified by \citet{L79b},
with a reported 5\arcsec separation and magnitude difference $\Delta$R
= 1.2 mag from the primary. \citet{CR02} also noted the existence of
the M5.0 dwarf companion while estimating the distance at
12.6$\pm$0.7pc based on the spectral type. Our estimated photometric
distance placed the star at 12.6$\pm$3.7pc \citep{L05}. Our astrometric
parallax for the system however clearly places the star at the larger
distance of 17.51$\pm$0.4pc. The existence of an unresolved companion
explains the underestimate of the spectroscopic and photometric
distances.

\subsubsection{LSPM J0405+7116W-B}

This faint companion to G 221-27 is reported here for the first
time. At the 17.5pc estimated distance, the two component have a
projected distance of about 13AU, while both together have a projected
separation of 90AU from the primary component (G 221-27). Note that
our parallax estimate for this triple system is based on the
astrometric motion of the primary, which is cleanly resolved from the
two companions on all the frames and is thus not affected by the
presence of the two companions.

\subsubsection{LSPM J0439+1615 = LHS 1690}

This is one of the very high proper motion stars from the LHS catalog
\citep{L79a}; its high proper motion was confirmed and re-measured by
\citet{BSN02}, \citet{SG03}, and \citet{LS05}. It was identified as a
probable nearby star by \citet{CR02} based on a spectroscopic distance
estimate of 12.3$\pm$1.1pc, while our photometric estimate placed the
star at 12.0$\pm$4.0pc \citep{L05}. Both estimate are largely
consistent with our astrometric parallax, which places the star at
$11.55\pm0.33$pc. We classify the star an M5.5e dwarf, with a weak but
clearly detected H$\alpha$ emission. The large proper motion yields a
transverse velocity of 43.7$\pm$1.2 km s$^{-1}$.

\subsubsection{LSPM J0510+2714}

This one was identified as a high proper motion star by \citet{LSR02},
confirmed by \citet{reid03}. It is a low galactic latitude object
(b=-7.4) in a relatively dense field. The star was identified as a
probable nearby star by \citet{R04}, based on a photometric distance
modulus of 0.70$\pm0.48$ mag, which suggests a distance of
$\approx14$pc. Our own photometric estimate placed the star at
10.1$\pm$3.2pc \citep{L05}. Our geometric parallax places the star
just within the 10-parsec horizon, at a distance of
9.93$\pm$0.16pc. The proper motion yields a transverse velocity of
31.4$\pm$0.5 km s$^{-1}$. Our spectrum reveals the star to be an
ultra-cool dwarf with spectral type M8.0e. The star has a strong
H$\alpha$ line in emision and is also detected in X-ray which suggest
it may be relatively young.

\subsubsection{LSPM J0515+5911}

Like the preceding object, this star was discovered as a high proper
motion star in a low galactic latitude field by \citet{LSR02},
re-confirmed by \citet{Levine2005}. It was classified as as M7.0 dwarf
by \citet{LRS03} and estimated to be at a spectroscopic distance of
14$\pm$4pc. Our spectrum is consistent with spectral type M7.5, and
our photometric distance places the star at 13.4$\pm$4.5pc
\citep{L05}. Both distance estimates are consistent with our parallax
measurement placing the star at 15.22$\pm$0.30pc. The very large
proper motion yields a transverse velocity of 73.5$\pm$1.5 km
s$^{-1}$, which flags it as a probable old disk object.

\subsubsection{LSPM J0711+4329 = LHS 1901}

This is another star from the LHS catalog \citep{L79a}, with its large
proper motion confirmed by \citet{BSN02}, \citet{SG03}, and
\citet{LS05}. The star was identified as a probable very nearby star
by \citet{R03} from a spectroscopic distance estimate of
8.0$\pm$0.8pc. \citet{S05} estimated a spectroscopic distance of 9.3
pc, while our own photometric estimate placed the star at
12.1$\pm$3.6pc. Our spectrum is consistent with a spectral type
M7.0. Interestingly, adaptive optics observations have resolved LHS
1901 into a pair of near equal-magnitude components, with a mean
angular separation of 0.2\arcsec \citep{Mont06}. Our geometric
parallax yields a distance of 12.85$\pm$0.50pc, which confirms the
spectroscopic and photometric distance underestimation from the
unresolved binary. The proper motion yields a transverse velocity of
41.1$\pm$1.6 km s$^{-1}$.

\subsubsection{LSPM J1119+4641 = LHS 2395}

Another star from the LHS catalogue \citep{L79a}, the star also known
as LP 169-22 was on the photometric list of \citet{weis96}, but was
formally identified as a nearby star by \citet{GR97}, who
estimated a spectroscopic distance of 18.2pc, from a spectral type
M5.5. \citet{RC02} estimated a photometric distance of 10.7$\pm$0.8pc,
while our own estimated suggested a photometric distance of
11.2$\pm$3.8pc. Our parallax places the star just beyond the 10pc
horizon at 10.31$\pm$0.28pc, with the proper motion yielding a
transverse velocity of 33.1$\pm$0.9 km s$^{-1}$.

\subsubsection{LSPM J1314+1320 = NLTT 33370}

This is a star from the NLTT catalog \citep{L79b} which was
re-identified by \citet{LS05} and has received little attention so far
(like many of the stars in the NLTT catalog not listed in the LHS.)
\citet{L05} identified it as a probable nearby star based on a
9.7$\pm$3.0pc photometric distance estimate. Our parallax places the
star at a much larger distance of 16.39$\pm$0.75pc. Our spectrum
yields a spectral type M7.0e with a very strong H$\alpha$ line in
emission. The significant underestimate in the photometric distance
strongly suggests that the star may be an unresolved
double. This would have to be tested by adaptive optics observations
or radial velocity monitoring. Alternatively the star could be
overluminous due to extreme youth, which would also be consistent with
the strong H$\alpha$ line. The proper motion yields a transverse
velocity of 23.8$\pm$1.1 km s$^{-1}$, consistent with the young disk
population.

\subsubsection{LSPM J1428+1356 = LHS 2919}

Another star from the LHS catalog \citep{L79a}. Identified as a nearby
star by \citet{R03} from an estimated spectroscopic distance of
13.0$\pm$1.4pc. \citet{S05} quoted a spectroscopic distance of 13.1pc,
while we estimated the photometric distance at 10.7$\pm$3.6pc
\citep{L05}. We classify the star as an M7.5e dwarf with weak but
detectable H$\alpha$ in emission. Our parallax places the star at
12.08$\pm$0.60pc, consistent with all the estimates above. Transverse
velocity is 34.6$\pm$1.7 km s$^{-1}$.

\subsubsection{LSPM J1757+7042 = LP 44-162}

This ultra-cool dwarf (M7.5) was identified as a probable nearby star
by \citet{gizis00} based on its very red optical-to-infrared color,
with a photometric distance estimate of 11.7pc (spectral type M7.5). A
photometric distance of 12.5$\pm$1.2 was estimated by
\citet{Cruz2003}, with our own estimate at 12.6$\pm$4.2pc
\citet{L05}. \citet{S05} estimated a spectroscopic distance of
16.0pc. Our parallax places the star at 19.08$\pm$0.40, significantly
larger than the photometric distance estimates, and suggesting the
star may be an unresolved double. The star was examined with adaptive
optics, but no companion was found
\citet{siegler2003,siegler2005}. Radial velocity observations might
reveal the star to be a spectroscopic double.

\subsubsection{LSPM J1817+1328}

This high proper motion star was discovered by \citet{LSR02}, and
spectroscopically identified as a white dwarf by \citet{LRS03}. The
star is identified to be within the 20pc horizon by
\citet{holberg2008}, from the photometric distance of 15.6$\pm$2.5pc
estimated by \citet{subasavage2007}. Our parallax indeed places the WD
at 14.22$\pm$0.24pc. Our spectrum suggests the object is a very cool
hydrogen white dwarf, with spectral type DA 11. The large proper
motion yields a transverse velocity of 80.4$\pm$1.4 km s$^{-1}$,
consistent with the old disk population.

\subsubsection{LSPM J1826+0146 = NLTT 46476}

This high proper motion star from the NLTT catalog was identified as a
probable nearby star by \citet{L05}, from a photometric distance of
12.6$\pm$3.7pc. We classify the star as an M4.5 dwarf. Our parallax
places the star at 18.69$\pm$0.70pc, which yields a transverse
velocity of 30.6$\pm$1.2 km s$^{-1}$.

\subsubsection{LSPM J1839+2952 = LP 335-12}

It was identified as a nearby star by \citet{RC02}. Spectroscopic
distances of 12.7pc and 13.1pc were estimated by \citet{R03} and
\citet{S05}, respectively, from a spectral type M6.5. This is to be
compared with our own photometric distance of 12.6$\pm$4.2pc
\citet{L05}. Our astrometry places the star at 12.61$\pm$0.32pc, with
a transverse velocity of 13.9$\pm$0.4 km s$^{-1}$.

\subsubsection{LSPM J1840+7240 = LP 44-334}

Identified as a nearby star by \citet{R04} from a spectroscopic
distance of 12.5$\pm$1.4pc (spectral type M6.5), while we estimated a
photometric distance of 13.6$\pm$4.6pc \citet{L05}. Astrometry places
the star at 16.86$\pm$0.63pc, with a transverse velocity of
15.3$\pm$0.6 km s$^{-1}$.

\subsubsection{LSPM J1926+2426 = G 185-23}

Another star from the Lowell Proper Motion catalog \citep{GBT71}, it
was identified as a nearby star by \citet{R04} from a spectroscopic
distance of 17.1$\pm$2.9pc, while we estimated the photometric
distance at 12.2$\pm$3.6pc \citet{L05}. Our astrometry places it at
18.94$\pm$0.54pc, in line with the spectroscopic distance. Transverse
velocity is 18.8$\pm$0.5 km s$^{-1}$. Our spectrum yield a subtype
M4.5e, with H$\alpha$ clearly detected.

\subsubsection{LSPM J2325+1403 = LP 522-46}

This NLTT star was flagged as white dwarf based on its location in the
reduced proper motion diagram \citep{SG02}, and spectroscopically
confirmed as a cool hydrogen (subtype DA 10) white dwarf by
\citet{VK03,KV06}. The star also was flagged as a cool white dwarf in
the Sloan Digital Sky Survey \citet{kilic2006}, and identified as a
member of the nearby WD population by \citet{holberg2008} with a
photometric distance of 18.7$\pm$6.0pc. Our astrometry places it at
22.27$\pm$0.99pc, beyond the 20pc horizon for the local white
dwarfs. Transverse velocity is 38.6$\pm$1.7 km s$^{-1}$.

\begin{figure}
\epsscale{1.1}
\plotone{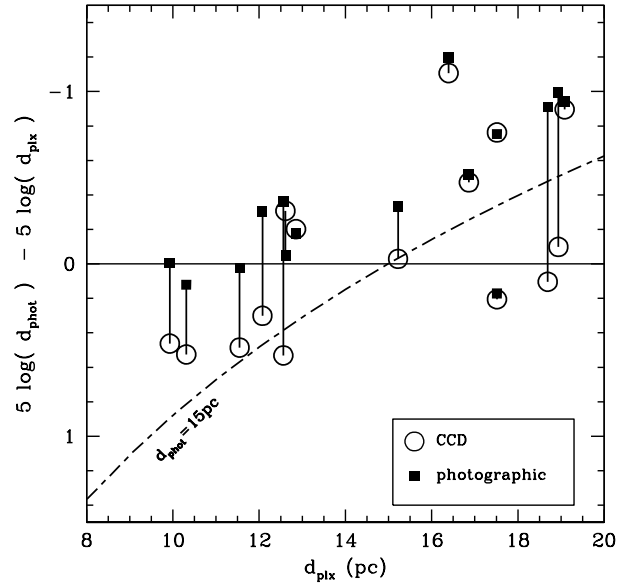}
\caption{Offsets between the photometric and parallax distances
  moduli, plotted against the parallax distance. Filled squares show the
  offsets for photometric distances moduli calculated using the
  photographic magnitudes $V_e$ quoted in the LSPM-north catalog. Open
  circles show the offsets for photometric distances moduli calculated
  using our CCD measurements. While the more accurate CCD photometry
  are a better predictor of the true distances for some stars, several
  stars remain with photometric distances significantly
  underestimating their true (parallactic) distances.}
\end{figure}


\section{Analysis and discussion}

One of our goal was to verify the reliability of the photometric
distance estimates for stars in the LSPM-north catalog, which is based
in part on photographic magnitudes. This is important for evaluating
whether more extensive lists of candidate nearby stars can be
efficiently assembled out of our proper motion catalogs. Current
photometric distances for stars in the LSPM-north catalog are based on
the [M$_{V_e}$,V$-$J] color magnitude relationship, calibrated by
\citet{L05} using nearby stars with trigonometric parallaxes. The
dispersion in the color-magnitude relationship ($\pm0.68$mag) suggests
that those distance estimates should be accurate to $\pm37\%$.

Figure 3 compares our photometric distance estimates to the measured
parallax distances. Filled squares plot the relative errors in the
distance moduli for $d_{phot}$ based on the photographic magnitudes
$V_e$. There is a clear trend for the more distant stars (d$>$14pc) to
have $d_{phot}$ underestimating their distances, while the nearer
objects (d$<$14pc) have their distances slightly overestimated by the
photometry. This can be explained by selection effects: our target
sample was largely assembled out of stars with $d_{phot}<15$ parsecs
(this selection selection limit is shown as a dashed line in
Fig.3). Overall, photographic distance moduli have a root mean square
(rms) error of 0.546 over the parallax distance moduli. These are
largely comparable with the rms for nearby stars with existing
parallax measurements \citet{L05} which show a 1-$\sigma$ dispersion
about the mean of 0.65.

To verify how much the errors $d_{phot}$ are due to
statistical/systematic errors on the photographic magnitudes $V_e$, we
also plot in Fig.3 the errors on $d_{phot}$ when calculated using the
photometric (CCD) magnitudes from our astrometric sequences (open
circles). We find that the CCD magnitudes do improve the accuracy of
$d_{phot}$ for two of the more distant stars (around 19pc), replacing
them in line with $d_{plx}$. However, the CCD magnitudes appear to
systematically overestimate the distances for the nearest
stars. Furthermore, three of the more distant objects stubbornly
remain with significantly underestimated distances. Overall, the CCD
photographic distance moduli have an rms error of 0.588, from which we
conclude that the CCD photometry does not significantly improve the
photographic distance estimates.

It is clear that significant errors remain which must be explained by
intrinsic uncertainties. Possible sources of bias on the photometric
distances include: (1) systematic or random errors in the photometry, (2)
unresolved binaries or multiples, which tends to underestimate the
distance of the pair, and (3) intrinsic scatter in the local
[M$_V$,V$-$J] color-magnitude relationship due to the dependence of
the color and absolute magnitude on age and metallicity. These biases
introduce uncertainties on the photometric distances of {\em
  individual} stars, no matter how well defined the mean
color-magnitude relationship is. The multiplicity fraction in M
dwarfs, in particular, is known to be $\approx40\%$ \citep{FM92}, so
significant numbers of doubles/multiple systems are expected to
contaminate lists of candidate nearby stars.

\begin{figure}
\epsscale{1.15}
\plotone{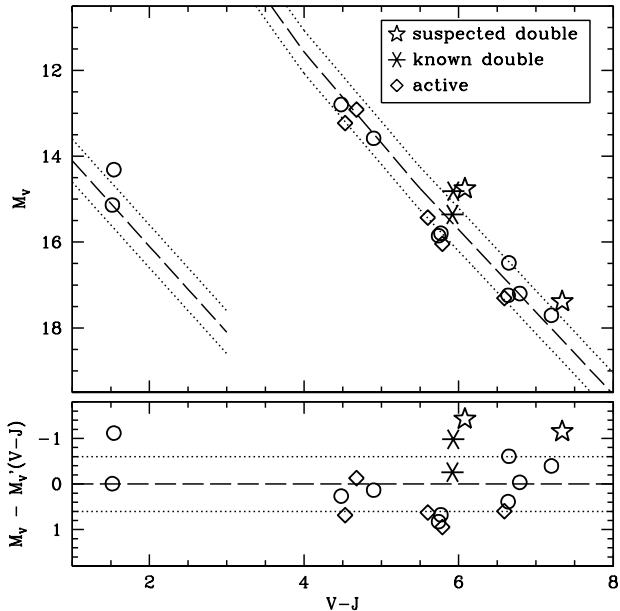}
\caption{Upper panel: color-magnitude diagram of our parallax targets,
  based the on CCD magnitudes and parallax measurements. The mean
  locus of the main sequence and white dwarf sequence, used to
  estimate photometric distances, are shown for comparison (dashed
  lines). The two stars denoted by star symbols (LSPM
  J1314+1320 and LSPM J1757+7042), hover significantly above the main
  sequence and are strongly suspected to be unresolved binaries. The
  known binary system LSPM J0405+7116E-AB, plotted as an asterisk,
  also falls significantly above the mean main sequence, as
  expected. Lower panel: residuals between the photometric and
  astrometric distance moduli, as a function of color. The dotted
  lines shows the intrinsic dispersion in the main sequence locus as
  observed by \citet{L05}. Our parallax targets show a similar
  scatter about the mean. No significant correlation is found between
  the V-J color and the residuals, which suggests that most of the
  dispersion is due to the intrinsic scatter.}
\end{figure}

We use our parallax measurements and improved photometry to construct
a color-magnitude diagram (CMD) for the stars on our program. The CMD
is shown in Figure 4. Most M dwarfs fall neatly along the main
sequence, while the two white dwarfs lie close to their expected
locus. We find that three of our program M dwarfs hover more than 0.8
magnitudes above the mean main sequence. One of them is the known pair
LSPM J0405+7116E-AB, which we resolved in our astrometric images; in
Fig.4 the unresolved pair is plotted as an asterisk. The other two
objects, which both hover more than 1 mag above the mean main
sequence, are the stars LSPM J1314+1320 and LSPM J1757+7042, which we
now suspect to be unresolved doubles; these are plotted as
stars in Fig.4. This suggest a possible $\approx10\%$ contamination
of the nearby star sample from distant unresolved doubles. The other
known unresolved pair, LSPM J0711+4329 (=LHS 1901), also plotted as an
asterisk in Fig.4, only falls marginally above the mean
color-magnitude relationship, which shows that not all unresolved
pairs should appear as significantly over-luminous. 

The mean main sequence and white dwarf sequence, shown by the dashed
lines in Fig.4 are the relationships we used to estimate the
photometric distances. The bottom panel in Fig.4 shows the residuals
between the photometric and astrometric distance moduli, as a function
of color. There does not appear to be any significant correlation
between the color and residuals. 

\begin{figure}
\epsscale{1.2}
\plotone{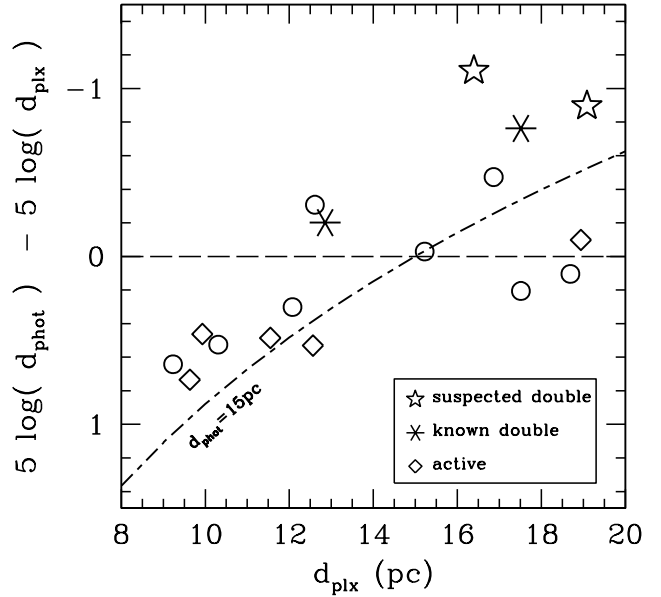}
\caption{Offsets between the photometric and parallactic distance
  moduli, with the photometric distances based on the ($V_e,V_e-J$)
  color-magnitude relationship of \citet{LS05}. The two suspected
  unresolved binaries, and the known pair LSPM J0405+7116E-AB are
  shown as open star symbols, as in Fig.5; all have photometric
  distances moduli underestimated by 0.7 mag or more. All the
  nearest stars ($d<12pc$) have their distances overestimated, when
  those distances are derived from photometry; we suspect selection
  effects to be responsible for the discrepancy (see text).}
\end{figure}

Figure 5 again shows the errors on the photometric distances
$d_{phot}$, calculated from the CCD magnitudes, similar to the bottom
panel in Fig.4 but now plotted as a function of distance. Different
symbols again are used for the known and suspected doubles, and also
for the more chromospherically active stars. Again no significant
difference is noted between the inactive and more active
objects, but there is a clear correlation between the residuals and
the parallax distances. This trend was already observed in
Fig.3, and appears to be related to selection effects. It is clear
that the more distant stars make it into the sample only if their
distances are underestimated by photometry. On the other hand, it
remains unclear why the more nearby stars should have their
photometric distances systematically overestimated. One possibility is
that this is due to selection effects from {\em earlier} parallax
programs. If these programs have been biased in favor of stars with
very short photometric distances, it is possible that the only nearby
stars left are those with overestimated photometric distances.


\section{Conclusions}

We have obtained reliable parallax distances for 18 high proper motion
stars identified in recent years to be probable nearby stars based on
photometric and spectroscopic distance estimates. Our parallax
measurements confirm that the stars are within the Solar Neighborhood,
with distances in the 9-22 parsecs range. Spectra are presented for
most of the stars, which confirms the white dwarf nature of two of the
targets.

Three of the stars are found to have absolute parallaxes placing them
within 10 parsecs of the Sun: PM J0011+5908 is an M6.5 dwarf at 9.2
parsecs; PM J0330+5413 is an active M4.5e dwarf at a distance of 9.6
parsecs; PM J0510+2714 is just within the 10 pc horizon, and is an
ultra-cool dwarf with subtype M8.0e. All three stars add to the
current census of $\approx300$ stars confirmed to be within 10 parsecs
of the Sun. The other 15 stars add to the tally of $\approx2,000$
confirmed systems with $d<25$pc.

Our sample illustrates some of the systematic and random errors that
plague photometric distance estimates. Even with accurate photometric
CCD measurements, several stars have photometric distance estimates
which significantly over- or underestimate their true distances. One
should thus expect some level of incompleteness and contamination in
photometrically selected samples of nearby stars, which indicates that
only systematic parallax measurements will produce reliable distances
of the now several thousand stars suspected to be in the 15pc-33pc
range \citep{L05}, and for which no geometric parallax measurement
exists to this date.

Our parallax measurements yield only a modest ($\simeq1\%$) increment in
the census of confirmed nearby stars. However, significantly larger
numbers of targets remain to be examined which promise to increase the
census significantly, especially beyond the 20pc horizon.

\acknowledgments

{\bf Acknowledgments}

This research was supported by National Science Foundation grant
AST-0607757 at the American Museum of Natural History. JT also
acknowledges support from the National Science Foundation through
grants AST-0307413 and AST-0708810.


\end{document}